\documentstyle[pra,preprint,aps]{revtex}

\begin{document}
\title{Existence of the Schmidt decomposition for tripartite systems}
\author{Arun Kumar Pati\footnote{~~email:akpati@sees.bangor.ac.uk}}
\address{School of Informatics, Dean Street, University of Wales, Bangor LL 57 1UT, UK.}

\maketitle

\tighten

\begin{abstract}
For any bipartite quantum system the Schmidt decomposition allows us to
express the state vector in terms of a single sum instead of double sums.
We show the existence of the Schmidt decomposition for tripartite system
if the partial inner product of a basis (belonging to
a Hilbert space of smaller dimension) with the state of the composite system
gives a disentangled basis. In this case the reduced density matrix of
each of the subsystem has equal spectrum in the Schmidt basis.
\end{abstract}

\vskip .5cm

\def\ra{\rangle}
\def\la{\langle}
\def\ver{\arrowvert}

\vskip 1cm

 The key to quantum information processing is the entangled nature of quantum
 states which has no classical counter part in the theory. These states are
 central to the study of quantum non-locality, quantum teleportation, quantum
 cryptography, dense coding and so on \cite{bd}. In the simplest term a
 pure entangled state is a
 one which cannot be expressed as a direct product of the states of two or
 more subsystems. If we have a quantum system consisting of two subsystems
 $A$ and $B$, then the state of the combined system in general can be
 expressed as $\ver \psi \ra_{AB} = \sum_{ij} a_{ij} \ver u_i \ra \otimes
 \ver v_j \ra$, where   $\{ \ver u_i \ra \}, (i=1,2,...N_A) \in {\cal H}_A$ and
 $\{ \ver v_i \ra \}, (i=1,2,...N_B) \in {\cal H}_B$ are the complete set of
 orthonormal basis vectors in their respective Hilbert spaces. The expansion
 coefficients in the above state of the combined system contain $N_A N_B$
 terms and is very difficult to manipulate with. However, the Schmidt
 decomposition (SD) theorem \cite{sch} comes to rescue
 us. It says that any arbitrary state of a bipartite (two-subsystem) quantum
 system can be expressed as \cite{ap}

 \begin{equation}
 \ver \psi \ra_{AB} = \sum_{i=1}^{N_A} \sqrt p_i \ver x_i \ra_A \otimes \ver y_i \ra_B,
 \end{equation}
 where $\{ \ver x_i \ra_A \}$ and $\{ \ver y_i \ra_B \}$ are two orthonormal
 basis sets belonging to Hilbert spaces ${\cal H}_A$ and ${\cal H}_B$,
 respectively and $N_A \le N_B$. The complex phases can be absorbed in the
 basis states and the expansion coefficients $\sqrt p_i$ can be
 chosen to be real and positive. This simplifies the expression
 to a great extent.

 The Schmidt decomposition theorem has been applied in many worlds
 interpretation of quantum theory \cite{eve}, in reproducing Clebsch-Gordon
 expansion of angular momentum states \cite{bcrs},
 in proving Bell's inequality
 \cite{ng,spdr} and is quite successful. Recently, in quantum optics context
 this has been applied \cite{aepk} and a geometric approach \cite{pka} to Schmidt
 decomposition of two-spin of particles is given in relation to Hardy's proof
 of quantum non-locality. However, all these discussions pertains strictly to
 bipartite systems only.

 If we have a composite system consisting of more than two subsystems there
 does not exist a Schmidt decomposition in general. But if it would exist, it
 will be quite useful simply because the number of terms one deals with in
 the expansion of the state vector will be comprehensibly small. For
 example, if we have a quantum system comprising of three subsystems (a
 tripartite system) then the general state of the system is given by

 \begin{equation}
 \ver \psi \ra_{ABC} = \sum_{ijk} a_{ijk} \ver u_i \ra_A \otimes
 \ver v_j  \ra_B \otimes \ver w_k \ra_C,
 \end{equation}
 where $\{ \ver u_i \ra \} \in {\cal H}_A= C^{N_A}$, $\{ \ver v_j \ra \}
 \in {\cal H}_B = C^{N_B}$  and $\{ \ver w_k \ra \} \in {\cal H}_C= C^{N_C}$.
 In this case there are $N_A N_B N_C$ terms to be dealt with. On the other hand if a
 Schmidt decomposition for tripartite system exists , then we can write the
 state in (2) as

 \begin{equation}
 \ver \psi \ra_{ABC} = \sum_{i} \sqrt p_i \ver x_i \ra_A \otimes \ver y_i \ra_B
 \otimes \ver z_i \ra_C,
 \end{equation}
 where $i = 1,2,... N_A$ (say) if $N_A = \dim{\cal H}_A$ is the smallest of
 all the three and 
 $\{ \ver x_i \ra_A \}, \{ \ver y_i \ra_B \}$ and $\ver w_i \ra_C$ are
 again orthonormal basis sets belonging to their respective Hilbert spaces. The
 general argument which goes against the existence of Schmidt  decomposition
 for a tripartite system such as (3) is that the ``equal-spectrum'' condition
 for reduced density matrices does not hold \cite{jp}.  Nevertheless, it is
 worth exploring under what conditions SD can exist. If the Schmidt decomposition
 for a tripartite system exists it would be useful in modal interpretation of
 quantum theory \cite{sk,ejs,rh}, for example. This possibility has been
 explored  by Peres \cite{per} and he found a necessary and sufficient condition
 for the existence of a Schmidt decomposition for a tripartite system.
 However, his condition does not give insight why does it fail for some
 tripartite systems and why does it {\it always work} for a bipartite system.
 Recently, Thapliyal \cite{at} has discussed Schmidt decomposability for a
 multipartite systems. He found
 that if the density matrix is multiseparable after tracing out any party,
 then the state is Schmidt decomposable.

 In this letter we find a simple criterion for the existence of Schmidt
 decomposition for tripartite system. This gives insight to the question: why
 does it work always for a bipartite not for a tripartite system. To state
 it simply, we prove that {\em if the partial inner product of a basis of any one
 of the subsystems (belonging to a Hilbert space of smaller dimension) with
 the state of the composite system gives a disentangled basis, then Schmidt
 decomposition for a tripartite system exists}. This condition turns out to be
 sufficient as well as necessary for the existence of SD. If the partial
 inner product gives an entangled basis the Schmidt decomposition in terms
 of a single sum does not exist, though the triple sum can be converted to a
 double sum. Using our existence criterion we show that the reduced density
 matrix of each
 subsystem  (by taking partial traces over any two subsystems) has the same
 eigenvalues, i.e. the ``equal-spectrum'' requirement holds. Our criterion
 is also consistent with the
 existence of Schmidt decomposition for a bipartite system. When we take
 the partial inner product of any one of the basis with the state of an
 arbitrary bipartite system, then the resulting basis belong to a Hilbert space
 of a single subsystem (no question of an entangled basis). This is the main
 reason why the Schmidt decomposition always works for a bipartite system.
 We also discuss the connection between our criterion and Thapliyal's
 multiseparability criterion for the existence of SD.

 Now we prove the following theorem.\\

 {\small {\bf Theorem}}: For any state $ \ver \psi \ra_{ABC} \in
 {\cal H}_A \otimes {\cal H}_B \otimes {\cal H}_C $ of a tripartite system let
 dim ${\cal H}_A = N_A$ is smallest of $N_B$ and $N_C$. If the ``partial inner
 product'' of the basis $\ver u_i \ra_A$ with the state $\ver \psi \ra_{ABC}$,
 i.e. ${_A\la} u_i \ver \psi \ra_{ABC} = \ver \psi_i \ra_{BC}$ has Schmidt number
 one then the Schmidt decomposition for a tripartite system exists.

 Proof: Let $ \ver \psi \ra_{ABC} = \sum_{ijk} a_{ijk} \ver u_i \ra_A \otimes
 \ver v_j \ra_B \otimes \ver w_k \ra_C$ and the partial inner product of
 the basis $\ver u_i \ra$ and state $\ver \psi \ra_{ABC}$ is a basis vector
 in the Hilbert space    ${\cal H}_B \otimes {\cal H}_C$ spanned by
 basis vectors
 $\{ \ver v_j \ra_B \otimes \ver w_k \ra_C \}$. It is given by

 \begin{equation}
 \ver \psi_i \ra_{BC} = \sum_{jk} a_{ijk} 
 \ver v_j  \ra_B \otimes \ver w_k \ra_C,
 \end{equation}
 where $\{ \ver \psi_i \ra_{BC} \}$ is an orthogonal basis  set but need not be
 normalised. We know that any vector in  a bipartite Hilbert space can be written
 as a Schmidt decomposition form, i.e.

 \begin{equation}
 \ver \psi_i \ra_{BC} = \sum_{\mu} \sqrt{ p_{\mu}^{(i)} }
 \ver y_{\mu}  \ra_B \otimes \ver z_{\mu} \ra_C,
 \end{equation}
 where  $\{ \ver y_{\mu} \ra_B \}$ and $\{ \ver z_{\mu} \ra_C \}$ are
 orthonormal basis for ${\cal H}_B$ and ${\cal H}_C$ which can be unitarily
 related to the basis $\{ \ver v_j \ra_B \}$ and $\{ \ver w_k \ra_C \}$,
 respectively. Therefore, the arbitrary state in general can be written as

 \begin{equation}
 \ver \psi \ra_{ABC} = \sum_{i \mu} \sqrt{ p_{\mu}^{(i)} }
 \ver u_i \ra_A \otimes \ver y_{\mu}  \ra_B \otimes \ver z_{\mu} \ra_C,
 \end{equation}

 Now there can be two situations: (i) either the basis (we call bi-Schmidt basis)
 $\{ \ver \psi_i \ra_{BC} \}$ is entangled or (ii) it is separable.
 Here, by entangled or separable basis we mean a basis all of whose members
 are so.  We can
 apply the pure state entanglement criterion, i.e., if the Schmidt number is
 equal to one then the state is separable and if it is greater than one it is
 entangled. Here, the Schmidt number is nothing but the number of non-zero
 eigenvalues in the reduced density matrix of a bipartite system and is the
 same
 as the number of terms in the Schmidt decomposition of a bipartite state.
 This is a good measure of entanglement for pure states. Now imposing this
 condition on bi-Schmidt basis, we write it as a separable form. So, if
 $\ver \psi_i \ra_{BC}$ has Schmidt number one we can write
 $\ver \psi_i \ra_{BC}  = \ver \beta_i \ra_{B} \otimes \ver \gamma_i \ra_{C}$.
 Since $\{ \ver \psi_i \ra_{BC} \}$ is not normalised 
 $\{ \ver \beta_i \ra_{B} \}$ and $\{ \ver \gamma_i \ra_{C} \}$ need not be
 orthonormal though they satisfy orthogonality condition.
 Therefore, the tripartite system can be written as

 \begin{equation}
 \ver \psi \ra_{ABC} = \sum_{i}  
 \ver u_i \ra_A \otimes \ver \beta_i  \ra_B \otimes \ver \gamma_i \ra_C.
 \end{equation}
 Now we calculate the reduced density matrix of each subsystem. The reduced
 density matrix for $A$ can be obtained by taking partial traces over $B$ and
 $C$. Thus,

 \begin{equation}
 \rho_A = {\rm tr}_B({\rm tr}_C(\rho_{ABC} )) = {\rm tr}_B [ \sum_{i}  
  q_i \ver u_i \ra_A {_A}\la u_i \ver \otimes \ver \beta_i \ra_B {_B}\la \beta_i \ver ],  
 \end{equation}
 where we have  used the trace equality ${\rm tr}_C( \ver \gamma_i \ra_C \la \gamma_j \ver) =
{_C}\la \gamma_j \ver \gamma_i \ra_C = q_i \delta_{ij}$ and  $q_i =
 \parallel \gamma_i \parallel^2 $ is the (squared) norm of the basis vector
 $\ver \gamma_i \ra_C$. Performing the second trace we can write the above
 one as

 \begin{equation}
 \rho_A =  \sum_{i}  
  q_i r_i \ver u_i \ra_A {_A}\la u_i \ver,
 \end{equation}
 where we have  used ${\rm tr}_B( \ver \beta_i \ra_B \la \beta_j \ver) =
 {_B}\la \beta_j \ver \beta_i \ra_B = r_i \delta_{ij}$ and  $r_i =
 \parallel \beta_i \parallel^2 $ is the (squared) norm of the basis vector
 $\ver \beta_i \ra$. Similarly, we can obtain the reduced density matrix
 $\rho_B$ and $\rho_C$ as

 \begin{eqnarray}
 &&\rho_B =  \sum_{i}  
  q_i r_i \ver \beta_i' \ra_B {_B}\la \beta_i' \ver   \nonumber\\
 && \rho_C =  \sum_{i}  
  q_i r_i \ver \gamma_i' \ra_C {_C}\la \gamma_i' \ver ,
 \end{eqnarray}
 where we have defined the orthonormal basis vectors $\ver \beta_i' \ra_B$
 and $\ver \gamma_i' \ra_C$ for ${\cal H}_B$ and
 ${\cal H}_C$ as $\ver \beta_i \ra_B = \sqrt r_i \ver \beta_i' \ra_B$ and
 $\ver \gamma_i \ra_C = \sqrt q_i \ver \gamma_i' \ra_C$. By comparing all
 the density matrices, we see that they have same eigenvalue spectrum
 $\{ q_i r_i \}$ in the Schmidt basis . Now we can redefine the state of the
 tripartite system as

 \begin{eqnarray}
 \ver \psi \ra_{ABC} & = & \sum_{i} \sqrt{ q_i r_i}
 \ver u_i \ra_A \otimes \ver \beta_i'  \ra_B \otimes \ver \gamma_i' \ra_C \nonumber\\
 & = & \sum_{i} \sqrt d_i \ver i \ra_A \ver i  \ra_B \ver i \ra_C,
 \end{eqnarray}
with $d_i = q_ir_i$ this is the Schmidt decomposition for a tripartite system
and hence the proof.

The above proof shows that the disentangled-partial-inner product crietrion
is a sufficient one. But it can be checked that it is also a necessary
condition, because if SD exists then the partial inner product of a basis
in the smallest Hilbert space with the joint state will give a disentangled
basis in other two tensor product Hilbert spaces.
 It should be remarked that $\rho_A, \rho_B$ and $\rho_C$ have the same
 number of distinct non-zero eigenvalues (non-degenerate spectrum), however,
 the number of zero-eigen values of  $\rho_A, \rho_B$ and $\rho_C$ can be
 different as  ${\cal H}_A , {\cal H}_B$ and ${\cal H}_C $ have different
 dimensions. The Schmidt decomposition (11) is unique for non-degenerate
 spectrum of reduced density matrices. The same is true for a bipartite
 system. Further, when we have a bipartite system then the ``partial inner
 product'' of any of the basis with the state of the system gives a single
 (disentangled) basis for the other one. Hence, the SD is always posible
 for a bipartite system.

 Once we know that the SD exists, then no local unitary operation of the form
 $U_B \otimes U_C$, local general measurements and classical communication
 can disprove the existence of Schmidt decomposition.
 We know that any measure of entanglement $E(\rho_{i,BC})$, with
 $\rho_{i,BC} = \ver
 \psi_i \ra_{BC} {_BC}\la \psi_i \ver $ satisfies the requirements \cite{pk}
 (i) $E(\rho_{i,BC}) = 0$ when $\rho_{i,BC}$ is separable,
 (ii) $E(\rho_{i,BC}) = E(U_B \otimes U_C \rho_{i,BC} U_B^{\dagger} \otimes
 U_C^{\dagger})$, and
 (iii) $E(\rho_{i,BC})$ cannot increase under local general measurement and
 classical communication, the Schmidt number of the bi-Schmidt
 basis does not change and it is impossible to disprove the existence of
 Schmidt decomposition.

 Next we discuss the situation when Schmidt decomposition does not exist
 for a tripartite system. This is the case when the bi-Schmidt basis is an
 entangled basis. Physically, this means there exists ``{\it entanglement within
 entanglement''}. When bi-Schmidt basis is separable there is entanglement
 between each subsystem $A, B$ and $C$ and there is no `` entanglement within
 entanglement''. When there is  ``entanglement within entanglement'' the
 ``equal-spectrum'' requirement breaks down. However, the ``equal-spectrum''
 holds within the subsystems $B$ and $C$, i.e., $\rho_B$ and $\rho_C$ have
 same non-zero eigenvalues. To see this consider the state of a tripartite
 system as  $\ver \psi \ra_{ABC} = \sum_{i}  \ver u_i \ra_A \otimes
 \ver \psi_i \ra_{BC}$. The density matrix  of the tripartite system is given
 by

 \begin{equation}
 \rho_{ABC} = \sum_{i j} 
 \ver u_i \ra_A {_A}\la u_j \ver \otimes \ver \psi_i \ra_{BC} {_{BC}}\la \psi_j \ver.
 \end{equation}
 On tracing over subsystem $B$ and $C$ we have the reduced density matrix $\rho_A$
 given by

 \begin{equation}
  \rho_{A} = \sum_{i} p_i
 \ver u_i \ra_A {_A}\la u_i \ver,
 \end{equation}
 where we have used the trace equality ${\rm tr}_{BC}(\ver \psi_i \ra_{BC}
{_{BC}}\la \psi_j \ver ) = {_{BC}\la} \psi_j \ver \psi_i \ra_{BC} = p_i \delta_{ij}$
 and $_{BC}\la \psi_i \ver \psi_i \ra_{BC} = \sum_k p_k^{(i)} = p_i$
 is the (squared) norm of the bi-Schmidt basis.
 The reduced density matrix $\rho_{AB}$
 given by

 \begin{equation}
 \rho_{AB} = {\rm tr}_C(\rho_{ABC}) = \sum_{ij \mu}  \sqrt{p_{\mu}^{(i)} p_{\mu}^{(j)}}
 \ver u_i \ra_A {_A}\la u_j \ver \otimes \ver y_{\mu} \ra_{B} {_B}\la y_{\mu} \ver.
 \end{equation}
 On tracing over the subsystem $A$ we get the reduced density matrix for
 $\rho_B$ given by

 \begin{equation}
 \rho_{B} =  \sum_{i \mu}  p_{\mu}^{(i)}  \ver y_{\mu} \ra_{B} \la y_{\mu} \ver
 = \sum_{\mu}  s_{\mu}  \ver y_{\mu} \ra_{B} {_B}\la y_{\mu} \ver,
 \end{equation}
 where we have defined $\sum_{i} p_{\mu}^{(i)} = s_{\mu}$ and each of them
 are some positive numbers. To obtain the reduced density matrix for
 $C$ we first note  that $\rho_{AC}$ is given by

 \begin{equation}
 \rho_{AC} = {\rm tr}_B(\rho_{ABC}) = \sum_{ij \mu}  \sqrt{p_{\mu}^{(i)} p_{\mu}^{(j)}}
 \ver u_i \ra_A {_A}\la u_j \ver \otimes \ver z_{\mu} \ra_{C} {_C}\la z_{\mu} \ver.
 \end{equation}
 From the above one we get the reduced density matrix for $\rho_C$ given by

 \begin{equation}
 \rho_{C} =  \sum_{i \mu}  p_{\mu}^{(i)}  \ver z_{\mu} \ra_{C} {_C}\la z_{\mu} \ver
 = \sum_{\mu}  s_{\mu}  \ver z_{\mu} \ra_{C} {_C}\la z_{\mu} \ver.
 \end{equation}
 From (15) and (17) these we can see that $\rho_B$ and $\rho_C$ have same eigenvalue
 spectrum $\{ s_{\mu} \}$ whereas the eigenvalue of $\rho_A$ has different
 spectrum $\{ p_i \}$. Thus, the ``equal-spectrum'' requirement breaks down
 when the bi-Schmidt basis $\{ \ver \psi_i \ra_{BC} \}$ has Schmidt number
 greater than one. Interestingly, if we look the subsystems $B$ and $C$ as a
 single subsystem $BC$, then the ``equal-spectrum'' requirement holds for
 subsystems $A$ and $BC$. This is because we have

 \begin{equation}
 \rho_{BC} =  \sum_{i}  \ver \psi_i \ra_{BC} {_{BC}}\la \psi_i \ver.
 \end{equation}
 On defining an orthonormal basis $\ver \psi_i' \ra_{BC}$ as
 $\ver \psi_i \ra_{BC} = \sqrt p_i \ver \psi_i' \ra_{BC}$, we have

 \begin{equation}
 \rho_{BC} =  \sum_{i} p_i \ver \psi_i' \ra_{BC} {_{BC}}\la \psi_i' \ver
 \end{equation}
This shows that $\rho_A$ and $\rho_{BC}$ have equal-spectrum as expected
 intuitively.

Before concluding we briefly discuss the link between our criterion
and Thapliyal's \cite{at}. We show that the multiseparability criterion discussed
in \cite{at} is a special case of our partial-inner product criterion when
each subsystem has equal dimension.
When we have a tripartite system and  all the Hilbert spaces
have equal dimension, then choosing a Hilbert space of lowest dimension
becomes degenerate. Therefore, one has to check if the partial-inner
product gives disentangled basis with respect to all the basis vectors and
all the parties concerned. For example, if
${_A\la} u_i \ver \psi \ra_{ABC} = \ver \psi_i \ra_{BC}$,
${_B\la} v_i \ver \psi \ra_{ABC} = \ver \phi_i \ra_{AC}$ and
${_C\la} w_i \ver \psi \ra_{ABC} = \ver \chi_i \ra_{AB}$ all gives
disentangled basis then there will be a Schmidt decomposition. However,
as can be seen if the above conditions hold the density matrices such as
$\rho_{BC}$, $\rho_{AC}$ and $\rho_{AB}$ are multiseparable.

In conclusion, we have found a simple criterion for the existence of Schmidt
decomposition for tripartite system and discussed why does it fail in
some cases. This also answers why does it always work for a bipartite system.
The existence of Schmidt decomposition might be useful in quantifying entanglement
content of a pure tripartite system. For example, if the SD exists then
the von Neumann entropy of
any of the reduced density matrix would give the entanglement content of
a pure tripartite system. Our criterion can be generalised to multipartite
(more than three) entangled states. Recently, generalised Schimdt-like
decomposition has been found for three qubit \cite{acin}
and multpartite cases \cite{tony}.

I thank S. L. Braunstein and A. Peres for going through my paper. I also thank
C. H. Bennett for useful remarks and interesting questions.
The financial support from EPSRC is gratefully acknowledged.

\end{document}